\documentclass[showpacs,prl,twocolumn,floatfix,
nofootinbib
]{revtex4}

\usepackage{amssymb}
\ifx\pdftexversion\undefined
  \usepackage[dvips]{graphicx}
\else
  \usepackage[pdftex]{graphicx}
\fi
\usepackage{bm}

\newcommand{\beq}{\begin{equation}}
\newcommand{\eeq}{\end{equation}}
\newcommand{\bea}{\begin{eqnarray}}
\newcommand{\eea}{\end{eqnarray}}

\newcommand{\pslash}[1]{\rlap{/}\kern-0.8pt #1}
\newcommand{\Dslash}{\rlap{/}\kern-2.0pt D}

\newcommand{\tr}{{\rm Tr}}

\def\today{\number\day\space\ifcase\month\or
January\or February\or March\or April\or May\or June\or
July\or August\or September\or October\or November\or December\fi
\space\number\year}
\newcount\mins \newcount\hours
\def\now{\hours=\time \mins=\time
	\divide\hours by60 \multiply\hours by60 \advance\mins by-\hours
	\divide\hours by60 
	\number\hours:\ifnum\mins<10 0\fi\number\mins}

\def\stacksymbols #1#2#3#4{\def\theguybelow{#2}
    \def\verticalposition{\lower#3pt}
    \def\spacingwithinsymbol{\baselineskip0pt\lineskip#4pt}
    \mathrel{\mathpalette\intermediary#1}}
\def\intermediary #1#2{\verticalposition\vbox{\spacingwithinsymbol
    \everycr={}\tabskip0pt
    \halign{$\mathsurround0pt#1\hfil##\hfil$\crcr#2\crcr
             \theguybelow\crcr}}}

\newcommand{\VEC}[1]{\mathbf{\bm{#1}}} 

\begin{document}


\title{Critical temperature for fermion pairing using
lattice field theory}

\author{Matthew Wingate}
\affiliation{Institute for Nuclear Theory, University of Washington,
Seattle, WA 98195-1550, USA}

\date{\today}


\begin{abstract}
Dilute gases of 2-component fermions are of great interest in 
atomic and nuclear physics.  When interactions are strong enough
so that a bound state is at threshold, universal behavior
is expected. 
Lattice field theory provides a first principles approach to
the study of strongly interacting systems such as this
through Monte Carlo simulation.  Results of 
exploratory simulations are presented here. 
In particular, the finite temperature phase transition between
superfluid and normal states is studied.  We present first results for
the critical temperature $T_c$ and describe the future work 
necessary to determine $T_c$ as a function of interaction
strength.
\end{abstract}


\maketitle


The ability to trap and cool bosonic atoms so they undergo Bose-Einstein
condensation (BEC) has created a unique route toward understanding
fundamental properties of quantum many-body systems.  Cooling fermionic
atoms to the same temperatures presents additional challenges due
to Pauli exclusion, but these are being overcome
\cite{Ohara:1999aaa,DeMarco:1999aaa}.

In contrast to BEC, in which a phase transition separates classical
from quantum behavior, the quantum nature
of Fermi gases appears gradually as the temperature reaches the Fermi 
temperature $T_F$.  A phase transition is expected for fermions
which have even the slightest attraction 
due to the Cooper instability, 
but this transition will occur far below $T_F$.
In the weak coupling limit, the system will form a
Bardeen, Cooper, and Schrieffer (BCS) superfluid below
the critical temperature which is exponentially small:
$T_c \propto T_F \exp(-\pi/(2 k_F|a|))$ \cite{Leggett:1980aaa},
where $a$ is the 2-body S wave scattering length 
The Fermi momentum and temperature are defined through the
number density $n$:  $k_F\equiv (6\pi^2 n/g)^{1/3}$ and $T_F\equiv k_F^2/2$;
$g$ is the degeneracy. (We will use the natural units $\hbar = M = k_B = 1$,
where $M$ is the fermion mass.)  Weak coupling implies $k_F|a|\ll 1$.

On the other hand, if the attraction between the fermions becomes
large, $a \to -\infty$ and a bound state appears at 
threshold.  As $1/a >0$ increases, the fermions pair to form
bosonic molecules which condense at a temperature which is
experimentally attainable, $T_c\approx 0.3 T_F$.  It is conjectured 
that $T_c$ will be close to $T_F$ even at $1/a=0$ \cite{Holland:2001}.
Much progress is being made toward producing and studying atomic
gases in this regime, 
\cite{Ohara:2002aab,Gehm:2003aaa,Bourdel:2003aaa,Chin:2004aaa}.

Furthermore, $1/a = 0$ is a special point. In this case the only
finite physical length scale is the mean interparticle spacing $n^{-1/3}$,
and the details of the interaction between fermions are irrelevant.
All physical quantities scale like appropriate powers of
$n$ times universal constants, applicable to atomic and nuclear systems.
Theoretical study of this limit is difficult because it occurs at
strong coupling.  

Numerical calculations have been done for this system at zero
temperature using fixed-node Greens function Monte Carlo
\cite{Carlson:2003prl,Chang:2004aaa}
and fixed-node diffusion Monte Carlo \cite{Astrakharchik:2004}.  
Within the context of the fixed node approximation and the
assumption that the initial trial wavefunction has a nonzero
overlap with the physical ground state, results for the
energy-per-particle and pairing gap were obtained.
These calculations are done at zero temperature with fixed
number of particles, up to 66 so far.

This work employs a very different Monte Carlo method based
on lattice field theory \cite{Chen:2003vy}, the same
approach widely used to compute
nonperturbative quantities in QCD \cite{Montvay:1994cy}.  
It has several advantages over previous methods.  There is no
sign problem in this formulation, so the fermions can be treated
exactly.  No assumptions need to be made regarding the ground state.
We work in the grand canonical ensemble, so the thermodynamic
limit is identically the infinite volume limit.  Lastly, the theory
is naturally formulated for finite temperature studies.  

Below we present the first numerical investigation of the fermion pairing
phase transition.  The method is described as pedagogically as
possible within the length constraints.  Computations are performed
along several paths in parameters space, and we observe a sharp decrease 
in the order parameter for fermion pairing as the temperature is 
increased.  These first calculations demonstrate the feasibility of using
lattice field theory to locate and compute the critical temperature
for fermion pairing.  Theoretical errors are discussed and
can be systematically removed.

We begin by considering the Hamiltonian for 2 species of 
equal mass particles, $N_1$ and $N_2$ in number:
\beq
H ~=~ -\frac{1}{2}\left(\sum_{i=1}^{N_1} \nabla_i^2
+ \sum_{j=1}^{N_2} \nabla_j^2 \right) 
+ \sum_{i=1}^{N_1}\sum_{j=1}^{N_2} 
v(|\VEC{r}_{i} - \VEC{r}_{j}|) .
\eeq
If scattering between particles is limited to
low energies, then the dominant contribution is the S wave.
Furthermore, if the potential vanishes quickly at long distances
an effective range expansion can be made for the scattering amplitude:
$(-1/a + \frac{1}{2}k^2 R + \ldots - ik)^{-1}$,
where $a$ is the scattering length and $R$ is the effective range of
the potential.  In a dilute gas $k R \approx n^{1/3} R \ll 1$, so 
scattering can be completely described by $a$; the short-distance
details of the potential are unimportant.  The potential can be
replaced by a local interaction, given in
second-quantized form by $\frac{1}{2}C_0(\bar\psi \psi)^2$,
where $\psi = (\psi_1,\psi_2)^T$ and $\bar\psi = (\bar\psi_1,
\bar\psi_2)$ are the fermion fields.
Negative $C_0$ corresponds to an attractive interaction.
Effective field theories have been developed which show how to 
systematically correct the short-distance physics should the 
need arise \cite{Kaplan:1998we,Braaten:2004rn}.  

We work in the grand canonical ensemble and rewrite the partition
function ${\cal Z} = \tr\exp [-\beta(\hat{H} - \mu \hat{N})]$ 
as a path integral
${\cal Z} ~=~ \int {\cal D}\psi {\cal D}\bar{\psi}~
\exp\left(-\int_0^\beta dx_4 \int d^3x ~{\cal L}\right)$.
The fermion fields have anti-periodic
boundary conditions in the imaginary time ($x_4$) direction.
In order to eliminate the Grassmann-valued fields, we must
make the Lagrange density quadratic in them.  To do so
an auxiliary scalar field $\phi$ is introduced 
(\`a la Hubbard-Stratonovich)
\begin{equation}
\frac{C_0}{2}\left(\bar{\psi} \psi \right)^2
\rightarrow \frac{1}{2}m^2\phi^2 \;-\; \phi\,\bar{\psi} \psi
\end{equation}
where $m^2 \equiv -C_0^{-1}$.

The Lagrangian is invariant under a global U(1) transformation
$\psi\to \exp(i\alpha)\psi$, $\bar{\psi}\to\bar{\psi}\exp(-i\alpha)$.
If we want to study the spontaneous breaking of this symmetry
in a finite volume, we must induce an explicit breaking and
compute observables in the limit of its removal.  Thus
we add to the Lagrangian a term proportional to a new parameter $J$:
\begin{equation}
{\cal L}[J] ~=~ {\cal L} \;+\; \frac{1}{2}\left( J\,\psi^T\sigma_2 \psi
\;+\; J^*\,\bar{\psi}\,\sigma_2\, \bar{\psi}^T \right) \, .
\label{eq:Lsrc}
\end{equation}
Without loss of generality we take $J$ to be real and define
the pairing condensate
$\Sigma \equiv \langle \psi^T\sigma_2 \psi
+ \bar{\psi}\sigma_2 \bar{\psi}^T\rangle/2$.  $\lim_{J\to 0}
(\lim_{\,V\to\infty} \Sigma)$ is an order parameter for the
breaking of the U(1) symmetry and the emergence of a nonzero
pairing condensate.  For the sign convention of
(\ref{eq:Lsrc}), $J>0$ will induce $\Sigma <0$.

${\cal L}$ is discretized on a lattice with 
$V$ spatial sites and $N_t$ sites in imaginary time; we denote
the spatial and temporal lattice spacings by $b_s$ and $b_t$.
To perform simulations, path integrals must be manipulated into
the following form \cite{Chen:2003vy}:
Combining $\psi$ and $\bar{\psi}$ into a 4-component
fermion, $\Psi^T \equiv (\psi^T, \bar{\psi}(i\sigma_2)^T)$,
the action can be written as $\Psi^T  {\cal A}  \Psi$, where
the $4VN_t\times 4VN_t$ matrix ${\cal A}$ is
\begin{equation}
{\cal A} \equiv \left(\begin{array}{cc}
-i J & K^\dagger \\ K & -iJ^*\end{array}\right) \left(\begin{array}{cc}
i\sigma_2 & 0 \\ 0 & i\sigma_2\end{array}\right) \, .
\label{eq:amatrix}
\end{equation}
$(K\psi)_x = (\psi_x - e^{\mu b_t}\psi_{x-\hat{e}_4}) - \frac{1}{2\xi}
(\VEC{\nabla}^2\psi)_x - \phi_x e^{\mu b_t} \psi_{x-\hat{e}_4}$,
and is the only $2VN_t\times 2VN_t$ block to contain nontrivial 
spacetime terms.
The only coupling between species is through the $i\sigma_2$ blocks.
We also introduce the anisotropy $\xi\equiv b_s^2/b_t$.
The result of the integration is
\beq
{\cal Z}  = \int {\cal D}\phi\, 
\det \left[ |J|^2 + \tilde{K}^\dagger\tilde{K}\right]
e^{-S_\phi} \equiv \int {\cal D}\phi \,W[\phi]
\label{eq:CKPI}
\eeq
where $K\equiv \tilde{K}\otimes I_2$ defines the $VN_t\times VN_t$ matrix
$\tilde{K}$.  The important point is that $W$, the integrand of ${\cal Z}$,
is strictly nonnegative, so it can be interpreted as a 
probability weight corresponding to a particular field configuration.


The generating functional for fermionic observables is obtained
by adding a fermionic source term $\mathfrak{Z}\Psi$ to ${\cal L}$
and integrating over $\Psi$.  The result is
\beq
{\cal Z}[\mathfrak{Z}] ~=~ \int {\cal D}\phi \, W[\phi]
\,\exp\left(-{1\over 2}\,\mathfrak{Z}^T {\cal A}^{-1}\mathfrak{Z}\right) \, .
\eeq
The expectation value of a general fermion bilinear is
generated by derivatives of ${\cal Z}[\mathfrak{Z}]$ as follows:
\beq
\langle \Psi^T {\cal B}\,\Psi\rangle = \left.
\frac{\delta}{\delta\mathfrak{Z}} \,{\cal B} \,
\frac{\delta}{\delta\mathfrak{Z}} \, {\cal Z}[\mathfrak{Z}]
\right|_{\mathfrak{Z}\to 0}
   =  -\mathrm{Tr}\, {\cal B\,A}^{-1} \, .
\label{eq:bilinear}
\eeq
The interesting bilinears in this work are the fermion density $n$ and the 
pairing condensate $\Sigma$, for which
\beq
{\cal B}_n = \left(\begin{array}{cc} 0 & i\sigma_2 S_+\\
i\sigma_2 S_- & 0
\end{array}\right)
~\mathrm{and}~~
{\cal B}_\Sigma = \left(\begin{array}{cc}
\sigma_2 & 0 \\ 0 & \sigma_2
\end{array}\right) 
\eeq
where $S_\pm$ denotes a temporal shift,
$S_\pm \psi_x =\psi_{x \pm \hat{e}_4}$.  
In practice a number of random sources $N_\mathrm{src}$
are used to evaluate the trace.  
Let $\eta_i(x,\alpha)$ be a Gaussian-distributed
random complex number at site $x$; $\alpha$ labels the 4 internal
$\Psi$ indices, and $i$ runs from 1 to $N_\mathrm{src}$.
Then $\mathrm{Tr} {\cal B\,A}^{-1} = \sum_i 
\eta_i^\dagger {\cal B\,A}^{-1} \eta_i /N_\mathrm{src}$.
The fermion density can
also be computed through $n = m^2\langle \phi\rangle$
$= m^2\int {\cal D}\phi \, \phi \, W[\phi]$; this provides a 
check of the code used to compute (\ref{eq:bilinear}).

As stated earlier,
this lattice field theory of a homogeneous, dilute gas
of 2-component fermions starts from
first principles, and uncertainties can be removed systematically.
The thermodynamic limit is reached simply by increasing the
volume of the system.  The continuum limit is more nuanced.
To remove the lattice spacing $b_s$ while maintaining constant
physics, one keeps $n^{1/3} a$ held fixed; then
all physical length scales are related to $n^{-1/3}$.  
The continuum limit is the limit where $n^{-1/3}\gg b_s$,
achieved by tuning lattice parameters so that 
$(b_s/a)\to 0$ and $\mu b_s^2/\xi \to 0$.

There are 3 physical quantities we wish to control: $n,a,$ and $T$.
We do so by tuning  $\mu, m^2$, and the size of the lattice
in imaginary time, varying either $N_t$ or $\xi$.  
A calculation in effective field theory matches the scattering length 
to the lattice parameters $m^2$ and $\xi$, modulo finite volume and 
spacing corrections \cite{Chen:2003vy}:
\beq
\frac{m^2}{\xi} = -\frac{b_s}{4\pi a} + \int_{BZ} \frac{d^3\VEC{p}}{(2\pi)^3}
\frac{1}{|\hat{\VEC{p}}|^2(1 + |\hat{\VEC{p}}|^2/4\xi)} \,.
\label{eq:matching}
\eeq 
$\hat{p}_i = 2\sin(p_i/2)$ and the $BZ$ integration/summation is
over available momenta in the Brillouin zone.
The temperature is tuned by varying $\xi$  
\beq
\frac{T}{n_0^{2/3}} = \frac{\xi}{N_t (n_0 b_s^3)^{2/3}} 
\label{eq:Tlatt}
\eeq
where $n_0$ denotes the zero temperature density.
One expects the chemical potential $\mu$ to strongly
control $n$.  However, the exact mapping of lattice parameters to 
physical quantities is not 
 is not known {\it a priori} and must be determined via simulation
before any precise physics results can be obtained.  This paper
is the first voyage into this parameter space.

The equivalence of the quantum statistical path integral (\ref{eq:CKPI})
to a classical statistical partition function suggests that 
numerical methods for simulating spin systems like the Ising model
can be used to study this dilute Fermi gas.  Additionally we can
draw from the expertise developed for lattice QCD.  For any
configuration of field values $C = \{\phi_x\}$, the contribution
of any observable $Q[\phi]$ is weighted by the nonnegative
factor $W[\phi]$.  One uses
importance sampling to generate an ensemble of 4 dimensional
field configurations which give the greatest contributions to 
$\langle Q\rangle = {\cal Z}^{-1}\int {\cal D}\phi \,Q[\phi] \, W[\phi]$.  
Having obtained a sample of $N$
configurations which are distributed according to $W[\phi]$, 
observables are estimated by their ensemble average $\overline{Q} =
\sum_C Q[C]/N$.

In this work, the
hybrid Monte Carlo (HMC) algorithm \cite{Duane:1987de} 
is used to generate samples of scalar field configurations.
One can think of the sequence of configurations as successive
snapshots of scalar fields evolving in an artificial time $\tau$,
referred to as Monte Carlo (MC) time (distinct from Euclidean
time $x_4$).  Given some initial configuration $C_0=C(\tau_0)$, a trial
configuration $C_0'(\tau_0+\Delta\tau)$ is obtained by evolving the 
4D scalar fields forward in $\tau$ using classical equations of motion.  
The process $C_0 \to C_0'$ is called a {\it trajectory} and is done
in discrete steps.  The 
change in the HMC Hamiltonian $\Delta {\cal H}$ is computed, and the new
configuration is accepted as $C_1$ with probability $\mathrm{min}
(1,e^{\Delta {\cal H}})$.
The effects of the fermions are included 
by rewriting the determinant as
a path integral over a complex scalar field $\chi$ with action
$\chi^* (|J|^2+\tilde{K}^\dagger \tilde{K})^{-1}\chi$.
The $\chi$ field is held fixed during the classical evolution.
At the end of a trajectory it is updated
by a heatbath step: $\chi = iJ^*\eta_1 + \tilde{K}^\dagger \eta_2$,
where $\eta_1,\eta_2$ are Gaussian noise vectors.

\begin{figure}[t]
\begin{center}
\includegraphics[width=8.0cm]{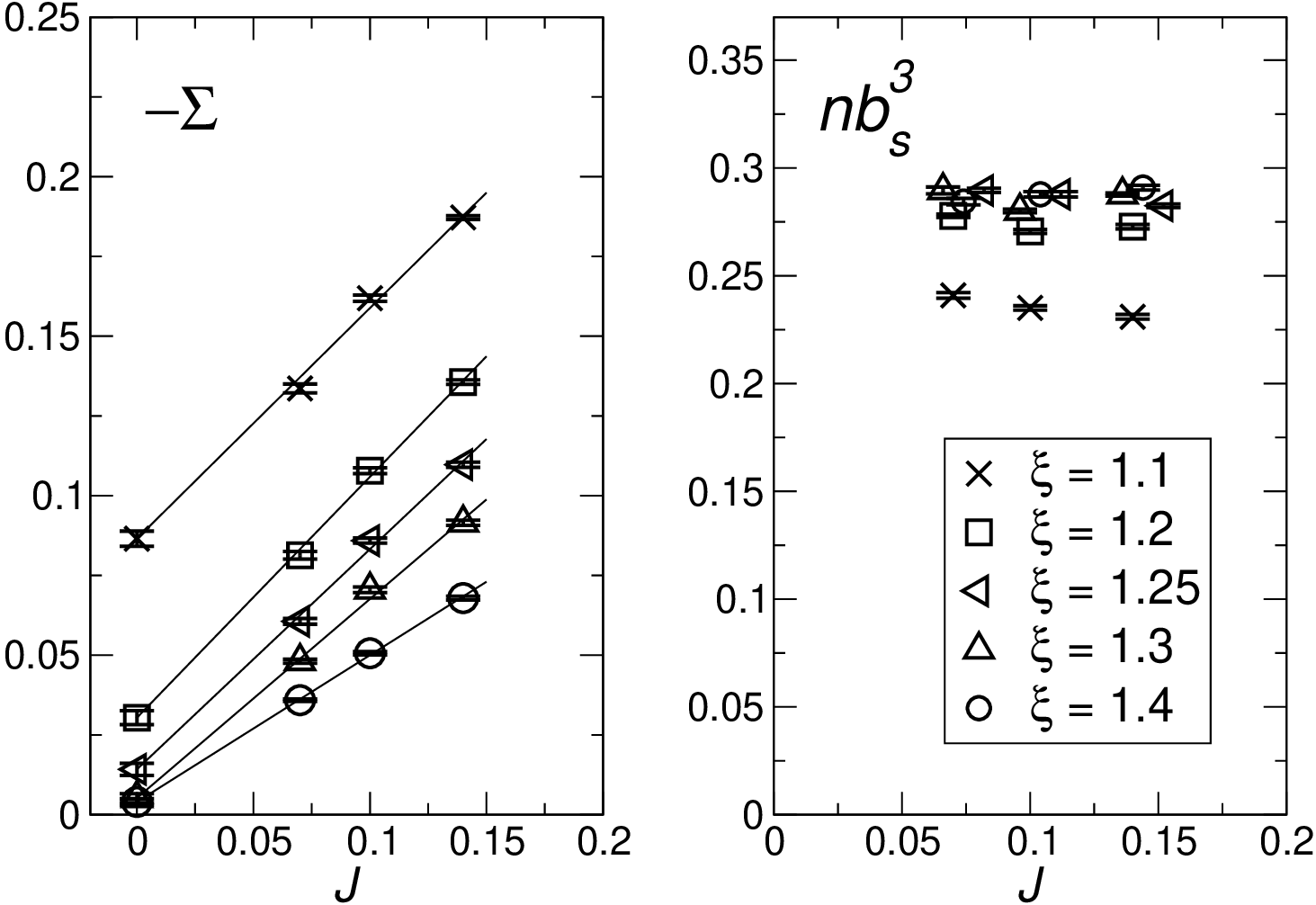}
\caption{\label{fig:SigN_mu0p4_m175} 
The fermion condensate $\Sigma$ (left) and the density $n$ in 
lattice units (right) as functions of external source for $m^2 = 0.175$.  
Different symbols indicate different values of $\xi$,
corresponding to different temperatures.  Points
at $J>0$ are simulation data, and those at $J=0$ are the results
of linear extrapolations (solid lines). Some data points are 
displaced horizontally for clarity.
}
\end{center}
\end{figure}

In some regions of parameter space, it is possible that the
attraction between fermions is so strong that the lattice is
densely packed with a fermion at every site, even at $\mu = 0$
\cite{Chen:2003vy}.
This lattice phase would be separated from the dilute, continuum-like
phase by a first order transition, preventing extrapolation to
zero lattice spacing.  Happily, it is straightforward to check 
that $n b_s^3 \ll 1$ for $\mu = 0$ 
in the interesting region of $(m^2, \xi)$ space,
namely where $b_s/a=0$ \cite{Wingate:2004wm}.
For example, on a lattice with $V=6^3$ and $N_t = 12$,
$nb_s^3 = 0.0073(6)$ when $\mu =0$ at $m^2=0.1456, \xi =1$ 
(corresponding to $1/a=0$ in infinite volume).  Due to finite
volume effects $n>0$.  The lesson is that these 
lattice parameters correspond to simulation of a dilute continuum-like 
Fermi gas, not a densely packed lattice of fermions.
Consequently, we can study thermodynamics at $\mu > 0$ confident
that the continuum limit can be reached in the manner described above.

The results presented below were obtained on a $V=8^3$ lattice
with $\mu b_t = 0.4$.
With $N_t = 16$, $m^2$ was set to 1 of 4 values, and then
4-5 values of $\xi$ were used to locate the phase transition.
At each $(m^2,\xi)$ pair, separate simulations were performed setting
$J = 0.07$, 0.1, and 0.14 in order to 
extrapolate to $J=0$.  Each simulation was run for 4000 HMC
trajectories with 32, 25, and 20 steps per trajectory 
for respective values of $J$.  Observables were computed 
(with $N_\mathrm{src} = 10$) at
20 trajectory intervals for a total of 200 measurements per simulation.
Dropping the first 50 measurements was sufficient to
ensure the sample had equilibrated.  To estimate correlations between
successive measurements, the data were binned in groups of 2, 5, 10,
and 20; $N_\mathrm{bin}=5$ sufficed to account for correlations.
The statistical errors are certainly much smaller than the
finite volume and lattice spacing uncertainties of this exploratory study, 
so we do not discuss them further.

\begin{figure}[t]
\begin{center}
\includegraphics[width=8.0cm]{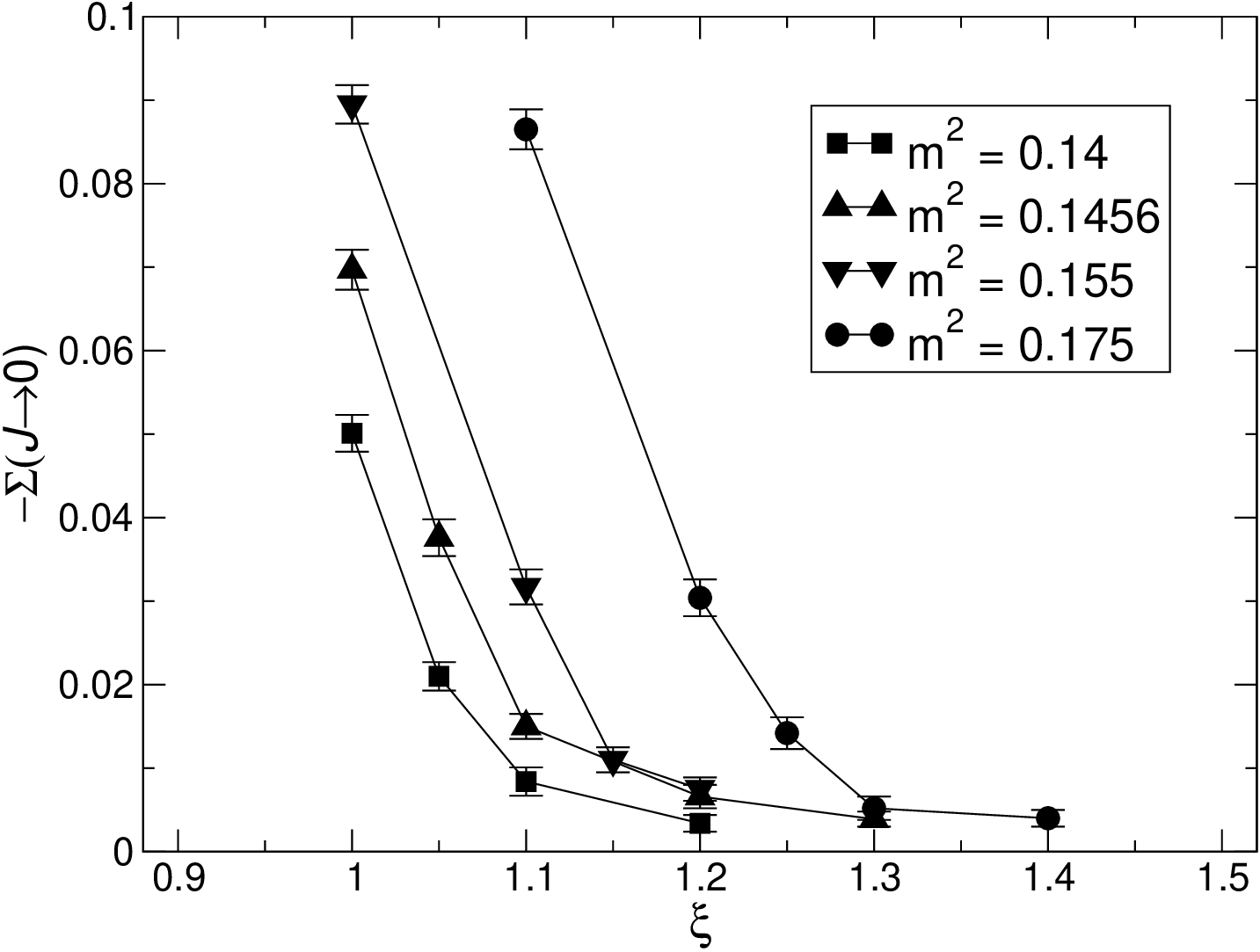}
\caption{\label{fig:orderplus_mu0p4_M} The fermion condensate $\Sigma$, 
after extrapolating $J\to 0$.  As $\xi$ increases, so does $T$,
and the system goes from superfluid to normal. Lines connect
the points merely to guide the eye.}
\end{center}
\end{figure}

Fig.~\ref{fig:SigN_mu0p4_m175} shows
the $J\to 0$ extrapolation of $\Sigma$ for $m^2 = 0.175$; 
the extrapolations for the other values of $m^2$ are similar.  
A more accurate extrapolation, exploiting an effective field 
theory analysis can be made once more data are obtained; however, 
in this work we emphasize the sharp decrease in $\Sigma|_{J\to 0}$.
Corrections to linearity will not qualitatively change our
exploratory conclusions about locating $T_c$.

In order to convert the observation of a phase transition at
a critical $\xi$ into a critical temperature using
(\ref{eq:Tlatt}), we need the
zero temperature fermion density $n_0$.
In this work we assume that $T_c/4$ is sufficiently low compared
to $T_c$ that we can compute $n_0$ with $N_t=48$ (and $J=0.14$).
We find $n_0 b_s^3 = 0.2612(6)$ and $0.2893(7)$ 
for $(m^2,\xi) = (0.155,1.2)$ and $(0.175,1.3)$, respectively.
(The quoted uncertainties represent the statistical error only.)
The densities at $T_c/4$ are close to those at $T_c$, e.g.\
the corresponding $n b_s^3 = 0.2577(12)$ and $0.2877(7)$.

Little dependence of $n$ on $N_t$ or $J$ 
is apparent in the finite temperature data.  Results for 
$n$ with $N_t = 16$ and $m^2 = 0.175$ are shown in
 Fig.~\ref{fig:SigN_mu0p4_m175};
plots for other values of $m^2$ are similar.

\begin{figure}[t]
\begin{center}
\includegraphics[width=8.0cm]{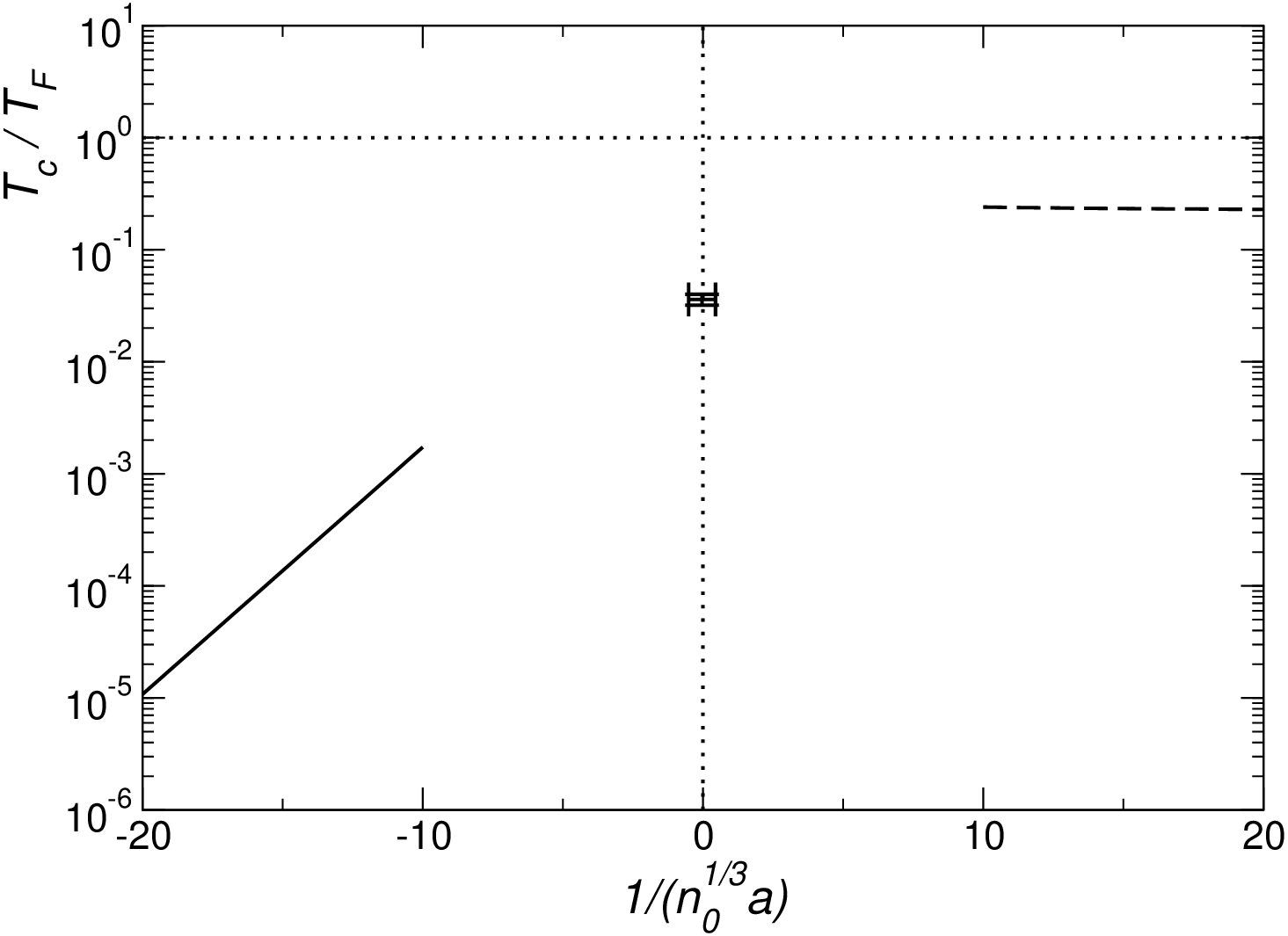}
\caption{\label{fig:Tc_ainv} Critical temperature vs.\
inverse scattering length. The solid line shows $T_c$
in the BCS regime and the bold broken line shows the
$T_c$ for BEC of di-fermion molecules.  
In the center is the present, {\it exploratory} result for $T_c/T_F$.
The horizontal error bar reflects the sensitivity of $a$ on $\xi$
over the critical region. 
}
\end{center}
\end{figure}

Using these low temperature densities in (\ref{eq:Tlatt})
leads to $T_c/T_F = 0.035 \pm 0.004$ and $0.036 \pm 0.004$
for $m^2=0.155$ and $0.175$.  
It turns out that $b_s/a$ changes significantly as
$\xi$ is tuned through the transition region.
With $m^2=0.155$, $b_s/a$ goes from $-0.10$ at $\xi=1.1$ to
$0.10$ at $\xi=1.2$, and with $m^2=0.175$,
$b_s/a$ varies from $-0.11$ at $\xi=1.2$ to $0.08$ at $\xi = 1.3$.
In order to determine $T_c$ precisely for
fixed values of $1/a$, simulations should hold $\xi$ fixed
and vary $\mu$ across the transition.  This work is in progress.


Fig.~\ref{fig:Tc_ainv} compares the results of this work 
with calculations of $T_c$ in the BCS 
\cite{Heiselberg:2000ya} 
and BEC
regimes.  
Since the different curves in Fig.~\ref{fig:orderplus_mu0p4_M}
correspond to the phase transition at slightly different lattice 
spacings (i.e.\ slightly different $n_0 b_s$) but over the same
large ranges of $1/(n_0^{1/3}a)$, only one data point is plotted.
Reduction of lattice spacing and volume errors will
allow $T_c$ to be studied more precisely in the region
$-1<n_0^{1/3}a<1$.

In conclusion, these initial results demonstrate both the
advantages and the challenges of using lattice field theory
to study fermion pairing at strong coupling. 
The critical temperature is straightforward to find;
we showed the condensate vanishing across the finite $T$
transition for 4 paths through parameter space.
The scattering length varies quickly along those paths,
so different directions in parameter space must be used to
compute $T_c$ while holding $a$ fixed.  The effects of lattice
volume and spacing need to be studying systematically and will
require significant computational resources.
This exploratory work represents the first leg of the journey
to obtain a first principles calculation of $T_c$ in the universal
regime.

This work was supported by DOE grant DE-FG02-00ER41132.
Simulations were performed on the LQCD clusters at
Fermilab and Jefferson Lab.
Conversations with A.\ Bulgac, J.-W.\ Chen, H.-W.\ Hammer, 
 D.B.\ Kaplan, and D.T.\ Son are gratefully acknowledged.

\bibliography{mbw}
\bibliographystyle{apsrev}


\end{document}